# The Research Based Learning - STEM Learning Activities: The Use of *r*-Dynamic Coloring to Improve the Students Metaliteracy in Solving a Tessellation Decoration Problem

A I Kristiana[1,2,3], Dafik[1,2,3], Z R Ridlo[3,4], R M Prihandini[2,3], R Adawiyah[2,3]

[1]Department of Maths Education Postgraduate, Universitas Jember, Indonesia
[2]Department of Maths Education, Universitas Jember, Indonesia
[3]CEREBEL, Universitas Jember, Indonesia
[4]Department of Sciences Education, Universitas Jember, Indonesia

E-mail: arikakristiana@unej.ac.id

**Abstract**: The metaliteracy ability is a very important in today's life, especially to live in the disruptive technology era. Metaliteracy is an ability that goes beyond metacognition and technological literacy. However, the students' metalliteracy ability is still relatively low. One of the causes of the low ability is due to the learning model that has been applied so far has not been able to bring out this ability. Therefore, in this study, an RBL learning model that is integrated with the STEM approach will be applied in solving the *r*-Dynamic coloring problem. By *r*-dynamic coloring, we mean a proper *k*-coloring *c* of *G* if for every vertex $v \in V(G)$ satisfies $|c(N(v))| \geq min\{r,d(v)\}$. The minimum *k* such that *G* has an *r*-dynamic coloring with k colors is called an r-dynamic chromatic number, denoted by *r(G)*. By using the *r*-dynamic coloring technique, we will improve students' metaliteracy in solving the tessellation decoration problem. Therefore, in this research, the syntax of learning activities of the Research-Based Learning and STEM approach will be developed including the assesment indicator of the metaliteracy ability.

INTRODUCTION

Metalliteracy is a comprehensive framework of thinking which includes information literacy (Jacobson, 2015). Metalliteration is divided into four domains, namely behavioral (what students should do after successfully completing learning activities and achieving skills and competencies), cognitive (what students should know after completing learning activities and achieving understanding, organizing, implementing, and evaluating), affective (is there a change in students' emotions or attitudes through involvement with learning activities), and metacognitive (what students know about their own thinking processes, reflective understanding of how they learn and solve problems, and how they can continue to learn). Metalliteration cannot be realized with the traditional classroom system, the use of the Internet of Things (IoT) and hypermedia in learning with a specific approach is one strategy to achieve metaliteracy. Schematically, metaliteracy indicators introduced by Jacobson and Mackey (2015) which have adopted the use of the Internet of Things can be presented in Table 1. There are five main indicators of metaliteracy, namely: produce (produce), incorporate (insert), use (use), share (divide). ), and collaborated (collaboration). These five indicators can be explained as follows:

Table 1. The indicators and sub-indicator of metaliteracy

| | |
|---|---|
| Produce | Identify the nature/characteristic of the problem; |
| | Develop breakthrough problem solving |
| | Develop or define stages, phases, syntax or algorithms. |
| Incorporate | Identifying patterns of solutions; |
| | Generalizing the solution; |
| | Using the Internet of Things, such as software, platforms and applications to integrate results. |
| Use | Test/assess results; |
| | Analyzing proven results |
| | Perform interpretation and prediction |
|  | Apply results. |
| Share | Using Internet of Things (Social Media, OER, MOOCs, Teaching Platform), Distribute the results followed by reflection and evaluation |
|  | Evaluation of feedback |
|  | Forecasting the response trend with software. |
| Collaborate | Working together using the Internet of Things platform; |
|  | Ask for the product taking into account some suggestions from others; |
|  | Encourage people to do more to contribute findings; |
|  | Obtaining joint works and products for publication; |
|  | Determining future work together for the wider community. |

One of specefic model and approach that can endorse the arise of the students metaliteracy is a combination of RBL and STEM. RBL stands for Research-Based Learning sedangkan STEM adalah Science, Technology, Engineering and Mathematics. Furthermore, RBL requires a contextual and realistic problem and includes at least four scientific studies, namely science, technology, engineering, and mathematics. Therefore, it is necessary to combine RBL with the STEM (Science, Technology, Engineering, and Mathematics) approach so that the students metaliteracy can be developed. Some research regarding the application of RBL and STEM in the classroom, it can be found in (Baharin et al. 2018, Breiner et al. 2012, Leon et al. 2015, Sergis et al. 2017, Siregar et al. 2019, Siregar et al. 2019, Soros et al. 2019). By combining the RBL and STEM approach, in this paper we will study "The Research Based Learning - STEM Learning Activities: The Use of r-Dynamic Coloring to Iprove the Students Metaliteracy in Solving the Tessellation Decoration Problem".

The main objectives of this research are as follows: (1) describe the framework of the RBL model learning activities with the STEM approach in solving the tessellation decoration problem by using $r$-dynamic coloring of graph, (2) describing the framework for the development process of learning materials on RBL model with STEM approach in solving the tessellation decoration problem by using $r$-dynamic coloring of graph (4) describes how learning materials of RBL model with stem approach can improve the metaliteracy in solving the tessellation decoration problem by using $r$-dynamic coloring of graph.

METHOD

This type of research is a qualitative method. The research starts from collecting some literatures and reviews. From the results of the literature review, we develop frameworks related to four research objectives above. The framework for the process of developing the RBL-STEM learning activities, referred to ADDIE model of the research and development, namely Analyze, Design, Development, Implementation, and Evaluation (Branch 2009).

RESEARCH FINDINGS

*Framework of RBL-STEM Syntax*

In the following, we will present a framework for integrating the RBL learning model in the STEM approach in improving the metaliteracy in solving the tessellation decoration problem by using $r$-dynamic coloring of graph. The framework is developed based on the syntax proposed in (Gita *et al.* 2021). In the early stages of RBL syntax is posing problems arising from the research group open problems. We consider the problem on tessellation decoration for **block paving tessellation** as follows.

*The decoration of block paving tessellation must be effectively designed to have symmetrical and good patterns. It must be designed to preserve uniqueness of pattern and also the accuracy of the number of block paving required for special color and shape. By mean of r-Dynamic coloring of graph, we will solve the problem of block paving tessellation decoration effectively.*

**SCIENCE**
Prior to layout the block paving, we need to have a good block paving tessellation decoration. Otherwise, we will have problem with the beauty and uniqueness of decoration including the accuracy of the block paving required.

**TECHNOLOGY**
By mean of some online tessellation software, Graph drawing software and also some others GUI mathematical software, we will solve the problem and involve the class students to solve it

**ENGINEERING**
To have a good *decoration of the block paving tessellation*, we will utilize the concept of r-Dynamic coloring of graph

**MATHEMATICS**
Testing the effectiveness of the use of the r-Dynamic coloring for the block paving tessellation decoration under a complexity math analysis by using analytics, qualitative, and deductive techniques by mean the algorithms development in the form of mathematical functions

**Figure 1.The STEM problem in developing the block paving tessellation decoration**

In this study, we consider the uniqueness of pattern of the decoration of block paving tessellation and also the accuracy of the number of block paving required for special color and shape. Therefore the RBL-STEM model will undertake the following stages (1) fundamental problems related to the block paving tessellation problem, (2) obtain a breakthrough by using *r*-dynamic coloring of graph, (3) data collection related to the data type being abused, (4) develop the decoration of block paving tessellation by using the *r*-dynamic coloring of graph, (5) test the resulting block paving tessellation decoration (6) reporting the research results and observations of students' metaliteracy. In detail, the framework for this RBL-STEM integration can be seen in Figure 2.

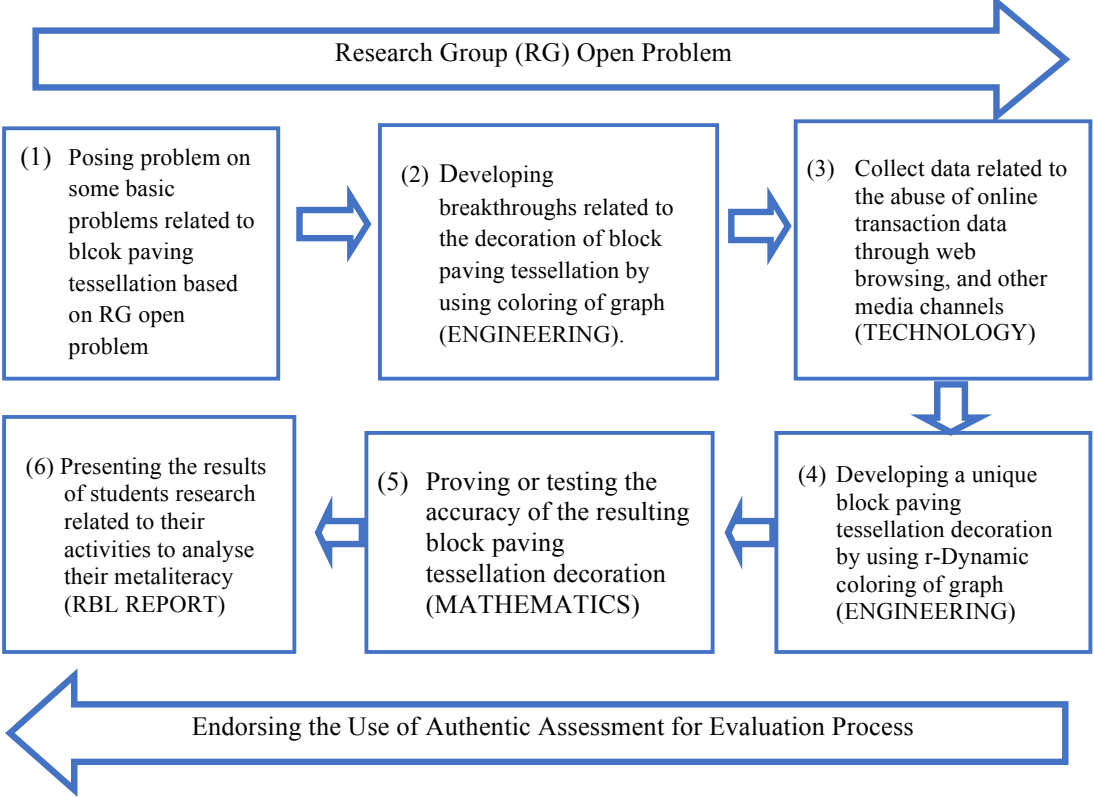

Figure 2.The framework of RBL-STEM in developing the block paving tessellation decoration

*Students Learning Outcome and Objective*

LEARNING OUTCOMES: Students can develop a tessellation decoration especially for block paving decoration by using *r*-dynamic coloring of graph. Students can also test whether the obtained using *r*-dynamic coloring of graph can be tested generally by using analytic and qualitative approach and the obtain the several possible decoration on block paving tessellation.

LEARNING OBJECTIVES: This RBL-STEM learning will enable students to develop knowledge and skills in the following fields of Science, Technology, Engineering, and Mathematics.

Sciences - Students are expected to:

- Understand the problem of the tessellation decoration problem especially on block paving decoration which is very useful on layouting the yard block paving process.
- Determine the application of the tessellation decoration problem in other use, especially for house wallpaper decoration.
- Analyse the decoration business strategy for gaining the maximum profit of business.

Technology - Students are expected to:

- Use a web browser to identify the concept of r-dynamic coloring and tessellation problems
- Use a researchgate site to find recent studies related to r-dynamic coloring and tessellation problems.
- Use the Youtube channel to find out the tutorial for tessellation maker by using Geogebra Software.
- Utilize the Geogebra Software for developing some various type of tessellation shape.

Engineering - Students are expected to:

- Develop a r-dynamic coloring of graph by using pattern recognition techniques
- Applying the r-dynamic coloring of graph algorithm in developing the tessellation decoration problem especially on block paving decoration

Mathematics - Students are expected to:

- Develop the r-dynamic coloring function by using the piecewise function technique
- Find the r-dynamic chromatic number by using an analytic and qualitative approach
- Use Matlab atau Excel software to develop a programming to calculate the number of block paving for special color required.

*On Block Paving Tessellation Decoration by Using r-Dynamic Coloring of Graph*

**The Element of Science Problem**. A tessellation or tiling of a flat surface is the covering of a plane using one or more geometric shapes, called *tiles*, with no overlaps and no gaps. Tessellations were used by the Sumerians (about 4000 BC) in building wall decorations formed by patterns of clay tiles. Now day, the most popular used of tessellation is for wall building and block paving decorations, see Figure 1 and Figure 2. The decoration of block paving tessellation must be effectively designed to have symmetrical and good patterns. It must be designed to preserve uniqueness of pattern and also the accuracy of the number of block paving required for special color and shape.

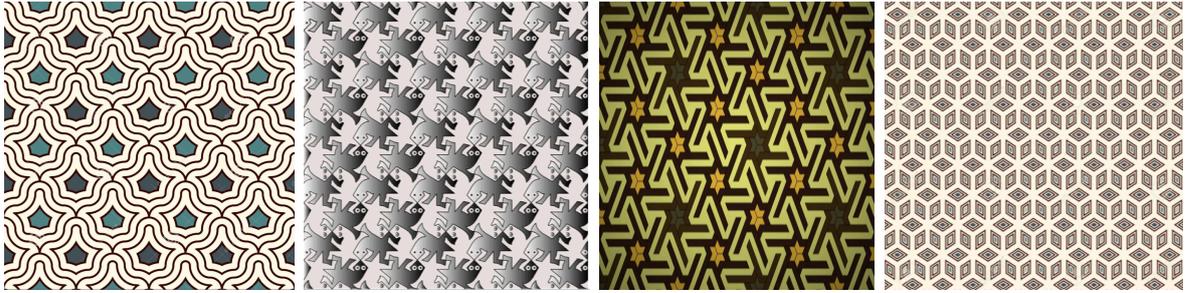

Figure 3. The used of Tessellation for building wallpaper decoration

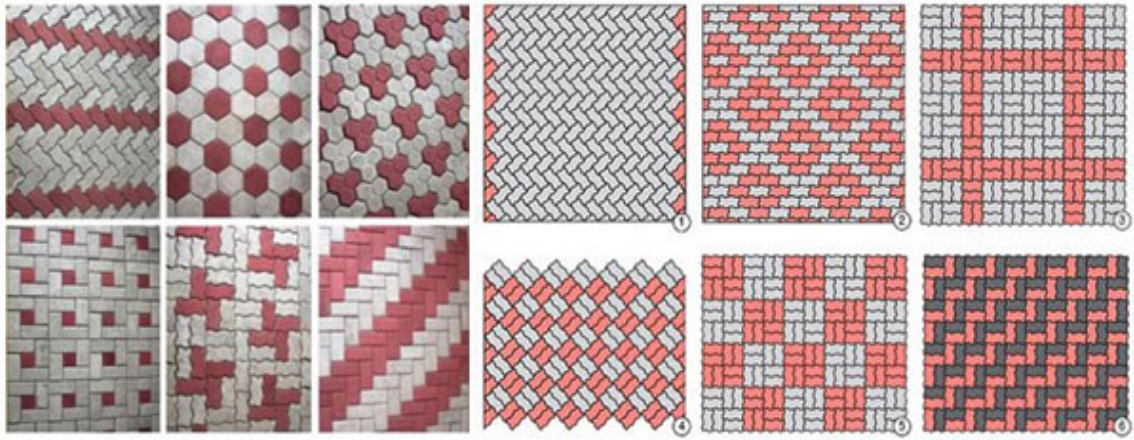

Figure 4. The used of tessellation technique for building block paving decoration

**The use of Internet of Things**. By mean of some online tessellation software, Graph drawing software and also some others GUI mathematical software such as Matlab. We will solve the problem and involve the class students to solve it. In this study we use Drawio online software (*visit https://app.diagrams.net*) to draw the tessellation of block paving decoration. For the illustration of using geogebra software for the tessellation maker, see Figure 3. This software can be accessed online. By choosing the GeoGebra Classic mode, and arrange the workspace by rightclick and set the axis and grid modes. Finally start to draw a polygon and combine with the pressing a point menu.

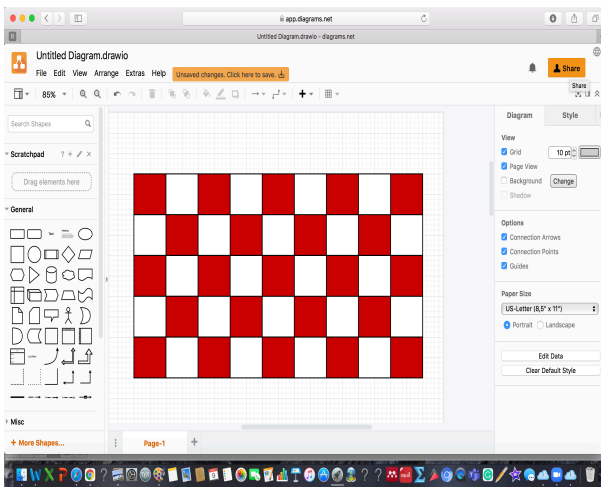

Figure 4. The illustration of tessellation sofware maker by using drawio online software.

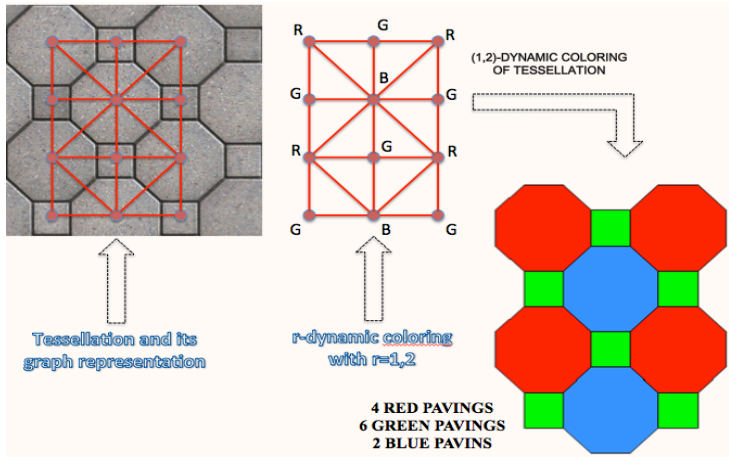

**The element of Engineering.** To decorate the block paving tessellation in terms of having a unique, symetrical and beautifull coloring, we need a special technique. The technique will be used in this study is a *r*-dynamic coloring technique. By an *r*-dynamic coloring of a graph *G*, we mean a proper *k*-coloring of graph *G* such that the neighbors of any vertex $v \epsilon G$ receive at least $min\{r, d(v)\}$ different colors, in other word we say $|c(N(v))| \geq min\{r, d(v)\}$. The *r*-dynamic chromatic number, written as $\chi_r(G)$, is the minimum *k* such that graph *G* has an *r*-dynamic *k*-coloring. To use the *r*-dynamic coloring, we need to do four steps, namely 1) determine the block paving tessellation as a contruction base, 2) draw its graph representation (a dual graph), 3) do *r*-dynamics coloring as the definition, 4) consider the obtained colors to decorate the block paving tessellation, 5) use the construction base of the block paving tessellation decoration to construct a biger tessellation complying with the area needed. Please make sure that there does not exist two adjacent block paving sharing the same colors.

Figure 5. The illustration of the use of r-dynamyc coloring in developing a tessellation

**The element of Mathematics.** Testing the effectiveness of the use of the *r*-Dynamic coloring for the block paving tessellation decoration under a complexity math analysis by using analytics, qualitative, and deductive techniques by mean the algorithms development in the form of mathematical functions and calculations. Figure 4. tells us for the *r*-dynamic coloring of *r=1*, and the *r*-dynamic chromatic number of this graph is 3. When we improve the parameter *r=2*, the number of colors does not sufficient, thus it needs to add one more colors. Thus $\chi_2(G)=4$. Furthermore for *r=2,* we need to find out the *3*-dynamic chromatic number $\chi_3(G)$ and finally the $\chi_r(G)$. To generalize this case, we need mathematical proof. Students are encouraged to involved in the proving process.

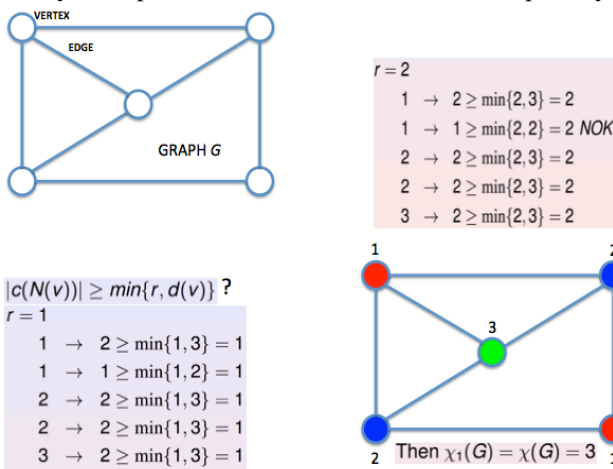

Figure 3. The illustration of the concept of *r*-dynamic coloring of graph.

In term of mathematics, we can define the simple graph representation above into the set of vertices $V(G)=\{x_1, ..., x_4, y_1, ..., y_4, z_1, ..., z_4\}$. The first permutation of *(1,2)*-dynamic coloring can be presented in the following function: $c(x_1, ..., x_4)$=*RGRG, $c(y_1, ..., y_4)$=GBGB, $c(z_1, ..., z_4)$=RGRG.* Furthemore, we can tabulate the following permutation tables.

**Tabel 2. The permutation of *(1,2)*-dynamic coloring of graph *G* with three colors**

| Color function of vertices | Color Permutations for *(1,2)*-dynamic coloring | | | | | |
|---|---|---|---|---|---|---|
| | (1) | (2) | (3) | (4) | | |
| $c(x_1, ..., x_4)$= | RGRG | GRGR | GBGB | BGBG | …. | …. |
| $c(y_1, ..., y_4)$= | GBGB | BGBG | RGRG | GRGR | …. | …. |

| $c(z_1, \ldots, z_4) =$ | | RGRG | GRGR | GBGB | BGBG | …. | …. |

**Tabel 3. The permutation of *3*-dynamic coloring of graph *G* with four colors**

| Color function of vertices | Color Permutations for *3*-dynamic coloring | | | | | | |
|---|---|---|---|---|---|---|---|
| | (1) | (2) | (3) | (4) | (5) | (6) | (7) |
| $c(x_1, \ldots, x_4) =$ | BYBR | BYBR | YBRB | YBRB | BYBR | YBRB | …. |
| $c(y_1, \ldots, y_4) =$ | RGRG | RGRG | GRGR | GRGR | YBYB | BYBY | …. |
| $c(z_1, \ldots, z_4) =$ | YBYB | BYBY | YBYB | BYBY | RGRG | RGRG | …. |

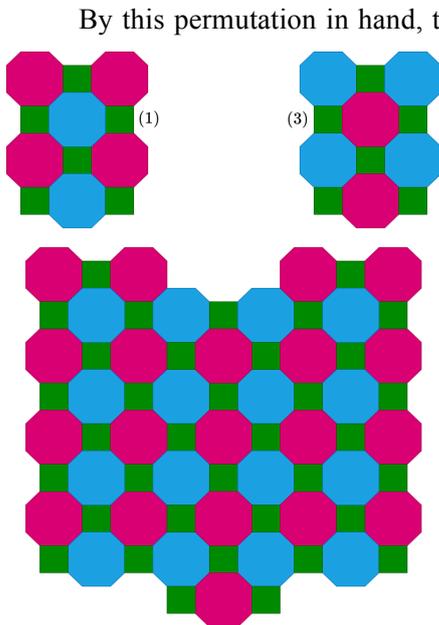

By this permutation in hand, then we can have the variation of block paving tessellation decoration in many decorations. When we choose the base construction of the two permutations of Table 1, namely *(1) and (3)*, we have the block paving tessellation decoration in Figure 5. It tells us that by mean of the *r*-dynamic coloring, it will be easy for seller and paving worker to construct the variation of decoration and count the number of block paving with special color. Consider if one base construction of 1 m$^2$ area. We need to decorate a yard of 200 m$^2$, then we need the block paving of red color as *4x100 + 2x100=600* pavings, green color as *4x200=800* pavings, blu color as *2x100 + 4x100=600* pavings. When we improve the valur of *r=3,* we need more colors to color all vertices. In this case, we need four colors, it implies the number of permutation also increase. Obviously, the seller can easily offer some choices to buyer for thei block paving tessellation decoration.

Figure 5. The illustration of block paving tessellation decoration by using *(1,2)*-dynamic coloring.

*RBL-STEM Learning Activities on Block Paving Tessellation Decoration*

In this section, we will discuss one by one the six stages of the RBL learning model complemented by STEM approach. These six stages will illustrate how students do in learning with RBL-STEM approach regarding the usse of r-dynamic coloring to improve the students metaliteracy in solving the tessellation decoration problem. Based on Figure 2, the first stage (SCIENCE) is proposing the fundamental problems related to the decoration of block paving tessellation. It must be effectively designed to have symmetrical and good patterns and the important one is the accuracy of the number of block paving required for special color and shape for yard block paving layouting. We will ask students to think about the block paving layouting in their yard. For more details, see Table 2.

**Table 4. The RBL-STEM learning activities on block paving tessellation decoration problem**

| STAGE ONE | LEARNING ACTIVITIES |
|---|---|
| (1) Posing problem on some basic problems related to block paving tessellation decoration (SCIENCE) | 1. The lecturer asks students whether they have ever seen the block paving tessellation process in their yard<br>2. The lecturer shows to the students a specific tessellation decoration and ask them wheather they can develop the block paving tessellation by them self or not.<br>3. The students works on developing a construction base with specific block paving tessellation decoration and try to analyse the required block paving for each color<br>4. The students start to have a group discussion |

The learning activities of RBL model with the STEM approach at the second stage (ENGINEERING) are developing breakthroughs related to the use of r-dynamic coloring to improve the students metaliteracy in solving the tessellation decoration problem. The Lecture asks students to identify some graph families who admit a r-dynamic coloring. For more details, see Table 3.

**Table 5. The RBL-STEM learning activities of r-dynamic coloring as a breakthrough**

| STAGE TWO | ACTIVITIES |
|---|---|
| (2) Developing breakthrough related to solve the tessellation decoration problem by using the $r$-dynamic coloring of graph (ENGINEERING) | 1. The lecturer guides students to discuss breakthroughs on how to solve the tessellation decoration problem by using the $r$-dynamic coloring of graph.<br>2. The lecturer explains to students how to decorate the block paving tessellation by using a simple graph representation.<br>3. The students are asked to find the $r$-dynamic coloring of simple graph on simple tessellation. |

The learning activities of the RBL-STEM approach on the third stage (TECHNOLOGY) are using online tessellation creator software, namely drawio application. See Table 4 for detail.

**Table 6. The RBL-STEM learning activities on online tessellation creator software usage**

| STAGE THREE | ACTIVITIES |
|---|---|
| (3) Utilizing the online tessellation creator software, namely Geogebra and Tesselation Maker software (TECHNOLOGY) | 1. Students under the guidance of lecturers download and try to use the online tessellation creator software, namely drawio.org.<br>2. Data collection related to $r$-dynamic coloring and tessellation shape is carried out by browsing the scientific journals/articles via research gates or other online library channels.<br>3. Students can use an encyclopedia, research gates (orcid, mendeley), research profile sites (scopus, publons), cloud storage (slideshare, Linkedin, MOOCs), cloud meetings (Google Meet, Zoom, Cisco Webex) to find the research results related to r-dynamic chromatic number of graph |

The learning activities by using the RBL model with the STEM approach at the fourth stage (ENGINEERING) is applying the $r$-dynamic coloring on block paving tessellation problems. This step begins by selecting the construction base of block paving tesselesation, and draw the graph representation and apply the obtained the $r$-dynamic coloring. For details, see Table 5.

**Table 7. The RBL-STEM learning activities on solving block paving tessellation by using *r*-dynamic coloring of graph**

| STAGE FOUR | ACTIVITIES |
|---|---|
| (4) Developing the block paving tessellation decoration by using the *r*-dynamic coloring of graph (ENGINEERING) | 1. Lecturers and students choose graphs for developing its *r*-dynamic coloring of graph and test its *r*-dynamic chromatic number.<br>2. Lecturers and students develop the block paving tessellation decoration by using the *r*-dynamic coloring with smallest *r*-dynamic chromatic number.<br>3. Lecturers and students select two possible base constructions with different permutation.<br>4. Lecturers and students try to generalize for bigger order of graph and develop the *r*-dynamic coloring of graph and test its *r*-dynamic chromatic number.<br>5. Construct the block paving tessellation decoration with obtained *r*-dynamic coloring. |

The learning activities by using the RBL model with the STEM approach at the fifth stage (MATHEMATICS) are proving that the *r*-dynamic coloring of any order and size of graph. Since the tessellation decoration may be in the form of large scale, thus the pattern availability should be determined by using mathematics concept. For more details, see Table 6.

**Table 8. The RBL-STEM learning activities on proving *r*-dynamic coloring of graph of any order and size and show an illustration for block paving tessellation decoration.**

| STAGE FIVE | ACTIVITIES |
|---|---|
| (5) Proving the accuracy of the resulting *r*-dynamic coloring of graph (MATHEMATICS) | 1. Define the sets of graph elements to determine the cardinality of the selected graph, i.e. triangular grid graph. Determine the number of vertices and edges to obtain the order and size<br>2. Determine the lower bound of the *r*-dynamic chromatic number based on the existing lemma<br>3. Determine the upper bound by developing the *r*-dynamic coloring function.<br>4. Evaluate the function whether the colors comply the properties of a proper *k*-coloring.<br>5. Does every vertex $v \in V(G)$ satisfies $|c(N(v))| \geq \min\{r, d(v)\}$<br>6. Compare the lower and upper bounds, if it is the same then we can set this number as the *r*-dynamic chromatic number<br>7. Take an example of block paving tessellation decoration of certain size of yard area to illustrate the use of its *r*-dynamic coloring by using a specific base construction. |

The learning activities by using the RBL model with the STEM approach at the six-stage (RBL REPORT) is a presentation of the research results related to the *r*-dynamic coloring to solve the block paving tessellation. In this case, students will take a focus group discussion (FGD), and the researcher can observe their metaliteracy. For more details, see Table 10.

**Table 10. The RBL-STEM learning activities on sharing the student's research results**

| STAGE SIX | ACTIVITIES |
|---|---|
| (6) Presenting the results of on students research related to the resulting key of r-dynamic coloring for triangular grid graph (RBL REPORT) | 1. Students develop a research report on the use of the *r*-dynamic coloring to solve the block paving tessellation.<br>2. Students do the presentation in front of the class to do focus group discussion and describe how does the techniques work especially for specific graph triangular grid graph.<br>3. Lecturers evaluate and clarify all the results of student's research activities.<br>4. Lecturers make observations on the students' metaliteracy by using observation sheets |

*The Instruments Framework for Assessing Students Combinatorial Thinking Skills*

The following will present the instruments framework of stduents **metaliteracy** assessment, see Table 11.

Sciences - Students are expected to:

- Understand the problem of the tessellation decoration problem especially on block paving decoration which is very useful on layouting the yard block paving process.
- Determine the application of the tessellation decoration problem in other use, especially for house wallpaper decoration.
- Analyse the decoration business strategy for gaining the maximum profit of business.

Technology - Students are expected to:

- Use a web browser to identify the concept of r-dynamic coloring and tessellation problems
- Use a researchgate site to find recent studies related to r-dynamic coloring and tessellation problems.
- Use the Youtube channel to find out the tutorial for tessellation maker by using Drawio software.
- Utilize the Drawio software for developing some various type of tessellation shape.

Engineering - Students are expected to:

- Develop a r-dynamic coloring of graph by using pattern recognition techniques
- Applying the r-dynamic coloring of graph algorithm in developing the tessellation decoration problem especially on block paving decoration

Mathematics - Students are expected to:

- Develop the r-dynamic coloring function by using the piecewise function technique
- Find the r-dynamic chromatic number by using an analytic and qualitative approach
- Use Matlab atau Excel software to develop a programming to calculate the number of block paving for special color required.

**Tabel 11. The framework of students metaliteracy assessment instruments**

| Indicator | Sub-Indicator | Test Framework |
|---|---|---|
| *Produce* | • Identify the properties/characteristic of the problems<br>• Obtain a breaktrough<br>• Develop the stages, phases, sintax or algorithm | - Identify the problem of the tessellation decoration problem<br>- Consider the block paving decoration for yard block paving construction.<br>- Determine the way how to decorate |

| | | the block paving tessellation<br>- Show the steps of the obtained technique for the block paving tessellation decoration. |
|---|---|---|
| *Incorporate* | • Identify the pattern of solution<br>• Develop generalization<br>• Use Internet of Things such as application or platform to integrate the results | - Identify the patern of the obtained technique for the block paving tessellation decoration.<br>- Expand the pattern to have a generalization.<br>- Identify the concept of *r*-dynamic coloring and apply it to have a base construction on the block paving tessellation decoration.<br>- Utilize the Drawio software Software for developing some various type of tessellation shape. |
| *Use* | • Test the results<br>• Analyse the results<br>• Interprete the results<br>• Implement the results | - Do simulation on the single base construction to develop wider block paving construction.<br>- Test whether the obtained r-Dynamic chromatic number of the base construction is optimum?<br>- Proceed the expantion with the mathematical proof.<br>- Implementing the result to develop yard block paving construction. |
| *Share* | • By using Internet of Things (Social Media, OER, MOOCs, Teaching Platform) circulate the results<br>• Have a reflection and evaluation toward a feedback<br>• Analyse the response to read the trend by using some application | - Develop a poster to share the result<br>- Write a a slide presentation or a paper to share the result in the scientific forum<br>- Develop a video to share the result in youtube chanel<br>- Have a feedback from the viewer and analyse the feedback for improving the result. |
| *Collaborate* | • Work with some other people by using IoT platforms<br>• Encourage someone to do more to contibutes the findings<br>• Determine a future work for wider society | - Follow up the feedback with requesting someone to collaborate for productive results.<br>- Elaborate some strategy to invite someone to do more for advacement<br>- Determine and forcast the future joint work |

*The Framework of Learning Material Process Development*

For this stage, we will incorporate a 4D model. It consists define, design, develop and dessiminate. The first: The **Analysis** stage is to analyze the characteristics of students, the material and learning process, and the learning media to be used. The second: The **Design** stage is to design the integration of RBL model into STEM approach. At this stage, the learning materials, namely syllabus, teaching learning plan, LKPD, pre-test, post-test, and other assessment instruments, are prepared by the researcher. The third: The **Development** stage is to test the learning materials and instruments to check the validity of the learning materials as well as the practicability. The results of the validation are in the form of content validity, format validity, and language validity, and the level of practicability. The fourth: The **Implementation** stage is to find the effectiveness of learning materials of RB-STEM in improving student's metaliteracy in developing the block paving tessellation decoration by using the r-dynamic coloring of graph.

The fifth: The **Evaluation** stage is a reflection activity to assess whether or not the application of RBL model learning materials with the STEM approach can improve students' metaliteracy in developing the block paving tessellation decoration by using the r-dynamic coloring of graph. In this stage, the use of inferential statistics is needed.

DISCUSSION

The development of a framework for STEM-RBL learning activities in developing the block paving tessellation decoration by using the r-dynamic coloring of graphs to improve students' metaliteracy is very useful study. This result will guide further researchers for doing an action research. There are at least two more research activities that can be done further, namely: (1) developing STEM-RBL learning materials with the ADDIE development model, (2) Studying the implementation of STEM-RBL learning materials in improving the students metaliteracy in developing the block paving tessellation decoration by using the r-dynamic coloring of graph. We have obtained that the framework of the learning activities of the combination RBL-STEM. It will be very effective in cultivating students metaliteracy which is in line with the previous research result presented in (Gita et al. 2020, Ridlo et al. 2020, Monalisa et al. 2019).

CONCLUSIONS

We have developed a framework of RBL-STEM learning activities in developing the block paving tessellation decoration by using the r-dynamic coloring of graphs to improve students' metaliteracy. This results as an initial activity to carry out further research activities, namely R & D and experimental research. However, this initial research is not easy, thus joint research for other STEM cases needs to be explored and as well as the breakthrough for solving the STEM problems are still widely open.

ACKNOWLEDGMENTS

We gratefully acknowledge the support from Combinatorial Education and Research-Based Learning (CEREBEL) of the year 2022.